\documentstyle[twocolumn,aps]{revtex}
\begin{document}
\draft              
\twocolumn[\hsize\textwidth\columnwidth\hsize\csname @twocolumnfalse\endcsname

\title{Single hole dispersion relation for the real CuO$_2$ plane.}

\author{V. I. Belinicher, A. L. Chernyshev, and V. A. Shubin}

\address{Institute of Semiconductor Physics, 630090, Novosibirsk, Russia}

\date{\today}
\maketitle

\begin{abstract}
Dispersion relation for the CuO$_2$ hole is calculated basing on the 
{\it generalized}\ \ $t$-$t^{\prime}$-$J$ model, recently derived from the 
three-band one. Numerical ranges for all model parameters, 
$t/J=2.4..2.7$, $t^{\prime}/t~=~0..-0.25$, 
$t^{\prime\prime}/t=0.1..0.15$, and three-site terms 
$2t_N\sim t_S\sim J/4$, have been strongly justified previously. Physical 
reasons for their values are also discussed. Self-consistent Born 
approximation is used for the calculation of the hole dispersion. 
An excellent agreement between calculated $E_{\bf k}$ and one 
obtained from the angle-resolved photoemission experiments is found.

\end{abstract}

]
\vskip 0.5cm 
%\twocolumn
\narrowtext

%************************************************************************

%************************************************************************

Recent angle-resolved photoemission experiments (ARPES) on insulating 
copper oxide Sr$_2$CuO$_2$Cl$_2$ \cite{Wells} can be considered as a 
direct test for low-energy models describing carriers (holes) in the 
CuO$_2$ plane.  Experimentally observed dispersion relation 
$E_{\bf k}$ for a {\it single} hole has the following 
characteristic features: (i) bandwidth about $2J$, (ii) band 
minimum at the $(\pi/2,\pi/2)$ point, (iii) isotropic dispersion near 
the band minimum, and (iv) almost flat dispersion along the line 
$(0,0) \rightarrow (\pi/2,0) \rightarrow (\pi,0)$.

The first result agrees with $t$-$J$ model prediction as well as with ones of 
all possible $t$-$J$ model generalizations in the region of 
parameters when t $>$ J. Following the physical arguments by Kane, 
Lee and Read \cite{KLR} it seems to be rather general that in 
presence of the strong spin fluctuations there are no stable 
quasiparticles at higher energies ($>2J$). It is of no importance 
whether the 'bare' dispersion of the hole exists 
($t$-$t^{\prime}$-$J$ model) or not ($t$-$J$ model), since the basic 
arguments are the absence of the hole--magnon scattering near the 
bottom of the band and its domination at higher energies.  Roughly, 
it looks like some kind of the Cherenkov effect: a massive 
quasiparticle cannot create an excitation with linear dispersion up 
to the threshold energy.

Experimental observation of the band minimum at $(\pi/2,\pi/2)$ point 
also agrees with  the quasiparticle (spin-polaron) dispersion 
calculated in the framework of $t$-$J$-like models.
It is well established by now that the almost degenerate dispersion 
along the magnetic Brillouin zone (MBZ) boundary ($(\pi,0)-(0,\pi)$ 
line) is an intrinsic property of the {\it pure} $t$-$J$ model, and that 
it is lifted out by any small (compared to $t$, not $J$) 
additional hopping integral, for example the 
next-nearest-neighbor hopping $t^{\prime}$.
Moreover, including $t^{\prime}$-term in the low-energy 
model of the real CuO$_2$ plane is strongly supported by the first 
principle calculations, which show that the direct O-O hopping 
provide large enough transfer amplitude to the next-nearest-neighbor 
site \cite{Saw}.

Thus, it is not much surprising that the masses in the directions 
along and perpendicular to the MBZ boundary were found close to each 
other.  To be considered as the experimental constraint on the 
parameters of the $t$-$t^{\prime}$-$J$ model at $J/t=0.4$ it fixes 
$t^{\prime}$ near $-0.3t$ \cite{Naz1}.

Returning to the experimental results, note that theoretical 
description of the last feature, i.e., flat band along 
 $(0,0) \rightarrow (\pi,0)$
line presents a problem. This flat region is absent in the $t$-$J$ 
model quasiparticle band. Simple $t^{\prime}$-term adds the 'bare' 
hole dispersion in the form: $\epsilon^0_{\bf k} = 4t^{\prime} 
\cos{k_x} \cos{k_y}$. Considering this term as the correction to the 
pure $t$-$J$ model dispersion one can see that it {\it does not} 
lift $(\pi/2,0)$ point from its $t$-$J$ model position, and so it  cannot 
provide the flat dispersion.

Disagreement between experimental band shape and theoretical one 
based on the $t$-$t^{\prime}$-$J$ model returns us to the problem of 
the correct low-energy model of the real CuO$_2$ plane. There were 
some recent works devoted to this problem, which consider CuO$_2$ 
holes in the framework of the three-band model in the strong-coupling 
limit \cite{star}, \cite{good}. These calculations reproduce the 
experimental band-structure  much better than the 
$t$-$t^{\prime}$-$J$ ones, but some of the fitting parameters 
differ from those proposed in the cluster analysis of the 
spectroscopic data and \cite{Saw} electronic structure works 
\cite{Hyb}.

From our point of view the experimental and {\it pure }
$t$-$t^{\prime}$-$J$ model discrepancy is the reason  to revise 
approximations were made in obtaining this model for the CuO$_2$ 
plane, not to deny it.

In our previous works \cite{Bel0} we developed ideas of the 
three-band model low-energy reduction, firstly proposed by Zhang and 
Rice \cite{Zh1}. We performed the consistent quantitative mapping of 
the initial model to the single-band one using Vannier-ortogonalized 
basis of the oxygen states and canonical transformation approach 
\cite{Bel2}.  It allowed us to obtain the low-energy generalized 
$t$-$t^{\prime}$-$J$ model and to calculate the ranges of its 
parameters for the real CuO$_2$  plane \cite{Bel1}.

Our general statement \cite{Bel1} is that there are physical reasons for 
including some other terms except 
$t^{\prime}$ one, namely hopping term to the next-next-nearest 
neighbors and the so-called 'three-site' terms, into low-energy model. 
We also should stress that the simple addition 
of  the $t^{\prime}$-term alone to the $t$-$J$ model is too naive to 
give the correct description of the subtle details of the hole 
spectrum.

In this paper we show that the {\it generalized}\ \ 
$t$-$t^{\prime}$-$J$ model with the set of parameters,  which 
presented in our recent work \cite{Bel1}, reproduces the experimental 
bandshape at all ${\bf k}$-points quite well.

Hamiltonian of the generalized $t$-$t^{\prime}$-$J$ 
model has the form \cite{Bel1}:
\begin{eqnarray}
\label{2}
&&H = H_{t-J} + H_{t^{\prime}}, \nonumber \\
&&H_{t-J}= t \sum_{\langle ij\rangle,\alpha}
\tilde{c}^{\dag}_{i,\alpha}\tilde{c}_{j,\alpha} + J \sum_{\langle ij\rangle}
{\bf S}_i{\bf S}_j  ,  \\
&&H_{t^{\prime}}=t^{\prime}\sum_{\langle ij 
\rangle_{2},\alpha}\tilde{c}^{\dag}_{i,\alpha}\tilde{c}_{j,\alpha}
+t^{\prime \prime}\sum_{\langle ij 
\rangle_{3},\alpha}\tilde{c}^{\dag}_{i,\alpha}\tilde{c}_{j,\alpha}
\nonumber\\
&&\phantom{H_{t^{\prime}}=} \mbox{}+t^{N}\sum_{\langle ilj 
\rangle,\alpha}\tilde{c}^{\dag}_{i,\alpha}\tilde{c}_{j,\alpha}\hat{N}_l
+t^{S}\sum_{\langle ilj 
\rangle,\alpha\beta}\tilde{c}^{\dag}_{i,\alpha} {\mbox{\boldmath 
$\sigma$}}_{\bar{\beta}\bar{\alpha}} \tilde{c}_{j,\beta} {\bf S}_l \nonumber
\end{eqnarray}
in standard notations of the constrained Fermi operators, brackets 
denote first ($\langle \rangle$), second and third ($\langle 
\rangle_{2,3}$) neighbor sites, respectively. Three-site terms are 
written in the rotationaly invariant form, $N_l$, ${\bf S}_l$  are 
the number of fermion and spin operators, respectively. $t_N$ and 
$t_S$ obtained for the usual Hubbard model are $2t_N=-t_S=t^2/U=J/4$, 
their 'three-band' values are not so simply related to the other 
parameters due to the presence of triplet state.

Ranges for the parameters of the model (\ref{2}) are \cite{Bel1}:
$t/J=2.4..2.7$, $t^{\prime}=0.01..-0.25$, $t^{\prime 
\prime}=0.12..0.16$, $t_N=0.01..0.07$, $t_S= -0.07..-0.16$. It 
is worth to note that the $t^{\prime}$ amplitude is smaller than it 
follows from the cluster calculation \cite{Saw} and $t^{\prime 
\prime}$ is not small compared to $t^{\prime}$. As it was discussed 
in Ref. \cite{Bel1} and in the work by Jefferson {\it et al} 
\cite{Jeff}, the main reason for the difference between Cu$_2$O$_8$ 
cluster and infinite plane $t^{\prime}$ hopping parameters is the 
Vannier nature of the latter. It was shown \cite{Bel1}, \cite{Jeff} 
that the Cu-O and O-O hopping amplitudes tend to cancel each other 
for $t^{\prime}$, and sum up for $t^{\prime \prime}$  -term.
This is the cause of not small and weakly waried $t^{\prime\prime}$, 
whereas $t^{\prime\prime}_{Cu-O}$ 
($t^{\prime\prime}_{O-O}$)$\ll$$t^{\prime}_{Cu-O}$ 
($t^{\prime\prime}_{O-O}$). 

Previously \cite{Bel1} we have calculated spin-polaron dispersion 
for the parameters discussed above using the simple variational 
ansatz \cite{Sush1}, which is quite good for the pure $t$-$J$ model. 
It consists of the 'bare' hole and four 'one-string' holes, that 
is, as it is clear by now, not enough for the correct treating of the 
$t^{\prime}$-terms.

In this paper we treat the energy calculation problem using the 
self-consistent Born approximation (SCBA). First of all, one should turn 
to the well known spinless-fermion Shwinger-boson representation for the 
Hubbard (constrained fermion) operators \cite{Schmit}. In that case 
constraint on the 
fermion degrees of freedom fulfilled exactly \cite{Horsch}, and the 
only 
approximation is the spin-wave one. Hamiltonian of the model (\ref{2}) 
becomes:
\begin{eqnarray}
\label{3}
&&H  \simeq  \sum_{\bf k}^{}\epsilon_{\bf k} h_{\bf k}^{\dag}h_{\bf k} +
\sum_{\bf q}^{}\omega_{\bf q} a_{\bf k}^{\dag}a_{\bf k} 
\nonumber\\
&&\phantom{H \simeq} \mbox{}+ \sum_{{\bf k},{\bf q}}^{}\left(M_{{\bf 
k},{\bf q}} h_{\bf k-q}^{\dag} h_{\bf k}a_{\bf q}^{\dag} + 
\mbox{H.c.}\right) + H^{(2)} 
\end{eqnarray}
where $h^{\dag} (h)$, $a^{\dag} (a)$, are the spinless hole and magnon 
operators, respectively, $\epsilon_{\bf k}$ is the 'bare' hole dispersion, 
$\omega_{\bf q}=2J(1-\gamma_{\bf q})^{1/2}$ is the spin-wave energy, 
$M_{{\bf k},{\bf q}} = 4 t (\gamma_{\bf k-q}U_{\bf q}+\gamma_{\bf k} 
V_{\bf q})$, $U_{\bf q}, V_{\bf q}$ are the Bogolubov canonical  
transformation parameters. $H^{(2)}$ includes the higher-order magnon 
terms. Bare hole dispersion has the form:
\begin{eqnarray}
\label{4}
&&\epsilon_{\bf k}=\epsilon_{\bf k}^0+ \delta \epsilon_{\bf k}
\nonumber\\
&&\epsilon_{\bf k}^0 = 4(t^{\prime} 
+ 2t_N- t_S)(\gamma_{\bf k}^2 - (\gamma_{\bf 
k}^-)^2) 
\\
&&\phantom{\epsilon_{\bf k}^0 =} \mbox{}+
 8(t^{\prime\prime} + t_N - t_S/2)(\gamma_{\bf k}^2 + 
(\gamma_{\bf k}^-)^2- 0.5) 
\nonumber 
\end{eqnarray}
where we used shorthand notations $\gamma_{\bf 
k}=\frac{1}{2}(\cos(k_x)+\cos(k_y))$, $\gamma_{\bf k}^- 
=\frac{1}{2}(\cos(k_x)-\cos(k_y))$.
$\delta \epsilon_{\bf k}$ is the addition from zero-point fluctuations:
\begin{eqnarray}
\label{5}
&&\delta \epsilon_{\bf k} = 4\alpha_1 (t^{\prime} 
+ 2t_N + t_S)(\gamma_{\bf k}^2 
- (\gamma_{\bf k}^-)^2) 
\nonumber \\
&&\phantom{\delta \epsilon_{\bf k}^0 =}\mbox{} + 8\alpha_2 (t^{\prime\prime}
+ t_N + t_S/2)(\gamma_{\bf k}^2 + (\gamma_{\bf k}^-)^2- 0.5) 
\\
&&\phantom{\delta 
\epsilon_{\bf k}^0 =}\mbox{}+ 4\beta t_S (4 \gamma_{\bf k}^2-1)
\nonumber 
\end{eqnarray}
where $\alpha_1=0.138$, $\alpha_2=0.107$, $\beta=-0.347$.
One can see from Eqs. (\ref{4}), (\ref{5}) that in contrast to 
$t^{\prime}$ (first terms), $t^{\prime\prime}$ (second terms) lift 
$(\pi/2,0)$ point to the higher energy.

Using SCBA we find the Green function of the hole as $G({\bf k}, \omega) = 
(\omega - 
\epsilon_{\bf k}- \Sigma({\bf k}, \omega -\omega_{\bf q}))^{-1}$ with 
the self-energy 
\begin{eqnarray}
\label{6}
\Sigma({\bf k}, \omega -\omega_{\bf q})= \frac{1}{N}\sum_{\bf q}  M_{{\bf 
k},{\bf q}}^2 G({\bf k-q}, \omega-\omega_{\bf q}) 
\end{eqnarray}
It was proved earlier \cite{Manos} that  the first order correction 
to the hole-magnon vertex is absent and the highest are very small.

Recently Bala, Oles and Zaanen \cite{zaan} showed that the 
higher-order vertices ($H^{(2)}$) do not change the SCBA results and 
confirmed that one-magnon coupling are accurate enough to reproduce 
the realistic properties of the $t$-$J$-like models. 

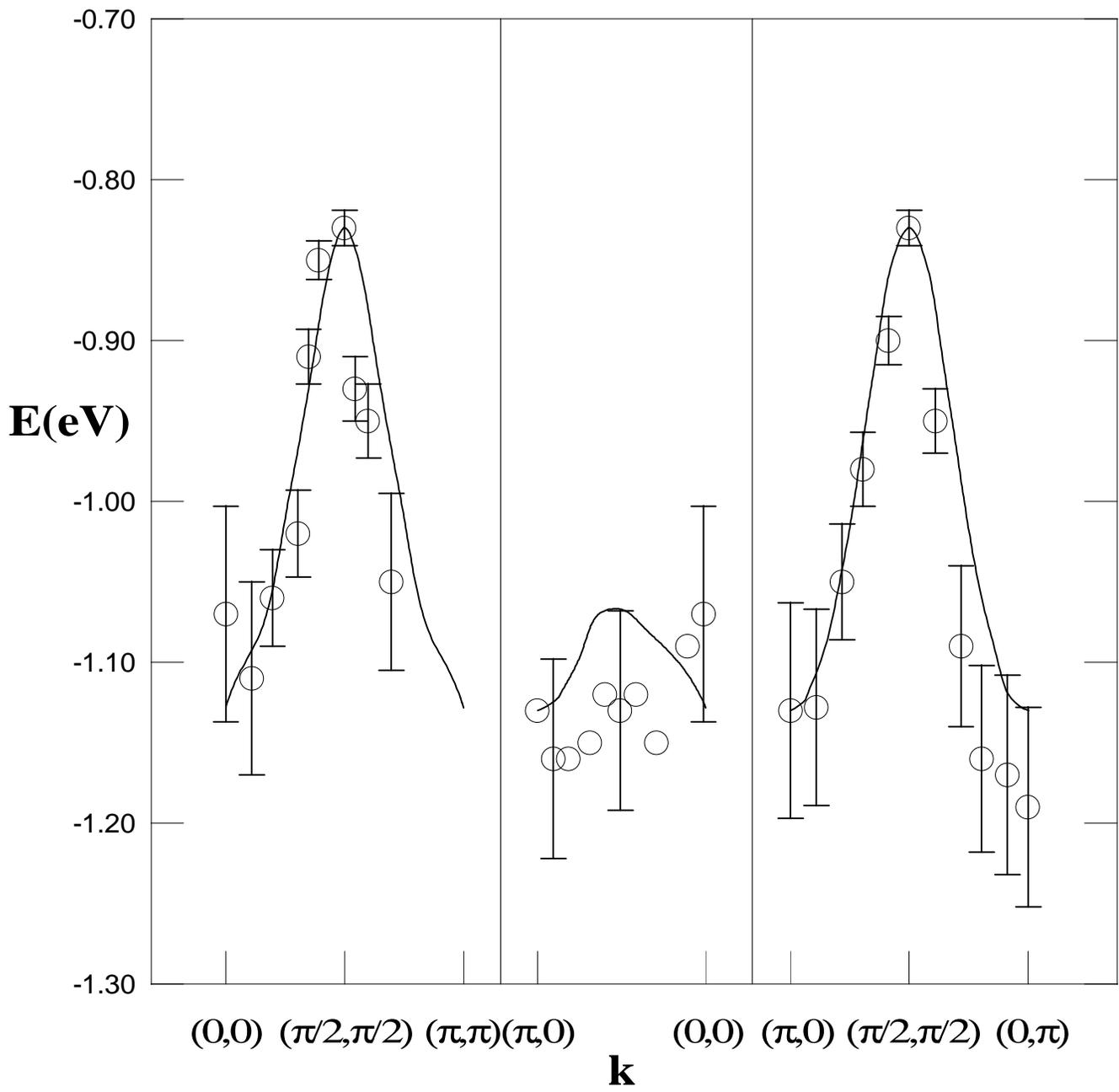
\begin{figure}
\caption{ 
Dispersion curve of a hole in the generalized  
$t$-$t^{\prime}$-$J$ 
model (\protect\ref{2}), (\protect\ref{3}) along the main 
directions $(0,0) \rightarrow (\pi,\pi)$, $(\pi,0)\rightarrow (0,0)$, 
and $(\pi,0) \rightarrow (0,\pi)$ (solid curve). Model parameters 
that provide this $E_{\bf k}$ are: $t/J=2.5, t^{\prime}=-0.2t, 
t^{\prime\prime}=0.15t, t_S=-2t_N=-J/4, J=0.14 $ eV. 
Experimental results from Ref. \protect\onlinecite {Wells} are 
also shown (open circles). 
} 
\label{fig1} 
\end{figure}

Equation (\ref{6}) was solved numerically  by the simple iteration 
procedure. We found no significant changes of the results for 
$16\times 16$ ${\bf k}$-points (in MBZ) and 1000 $\omega$-points, and 
for $24\times 24\times 3000$ points. Also, we checked our procedure 
for the pure $t$-$J$ model and found very close agreement with earlier 
results \cite{Dag}. Results of our generalized $t$-$t^{\prime}$-$J$ 
model calculations together with the experimental points are presented on 
Fig. \ref{fig1}. It is important to stress that it is {\it not}  
the 'best fit', we simply used the average values of parameters from 
their 'realistic range'.  In the main term of the bare dispersion  
$\epsilon_{\bf k}^0$ (\ref{4}) $t_N$ and $t_S$ terms enter in a 
combination $(2t_N-t_S)$, which realistic range is 
$[\frac{1}{4}J..\frac{3}{4}J]$, so we simply take its Hubbard value 
$(2t_N-t_S)=J/2=0.2 t$. We used  $t/J=2.5, t_N=J/8, t_S=-J/4$, and 
$J=0.14$ eV (from Ref. \cite{Hay1}). For $t^{\prime}$ and 
$t^{\prime\prime}$ we used $t^{\prime}=-0.2t$, 
$t^{\prime\prime}=0.15t$.  Note, that the bare dispersion (\ref{4}) 
consist of two terms, which can be considered as the 
$t^{\prime}_{eff}$ $(=t^{\prime}+2t_N-t_S)$ and $t^{\prime 
\prime}_{eff}$ $(=t^{\prime \prime}+t_N-t_S/2)$, and since 
$t^{\prime}$ and $(2t_N-t_S)$ have the opposite signs, 
$t^{\prime}_{eff}$ becomes very small at all realistic values of 
parameters.

Figure \ref{fig1} shows an excellent agreement with  experiment 
along $(\pi,0)-(0,\pi)$ as well as along $(0,0)-(\pi,0)$ lines. 
Notice, that another feature of the ARPES can be explained in the 
spin-polaron approach. Lower intensity of the photoemission peaks at 
 the top of the  hole band (bottom in the electron language) easily 
connected to the lower quasiparticle residue at $(0,0) \rightarrow 
(\pi,0)$ points. The last is due to importance of 
multi-magnon scattering processes for the 'cutting' of the wide 
initial ('bare') band \cite{zaan}.

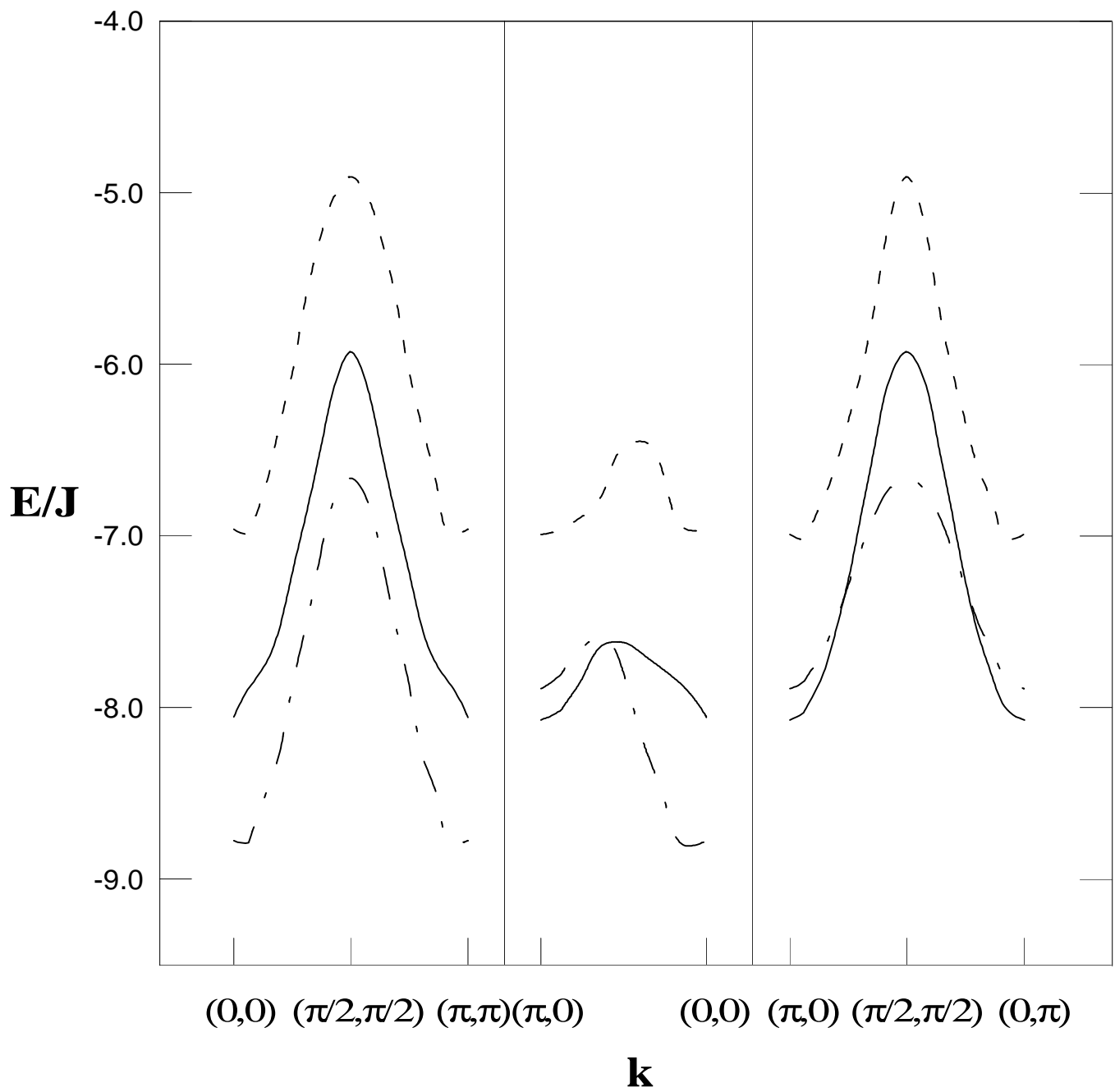
\begin{figure}
\caption{
Dispersion relation of a hole in the generalized   
$t$-$t^{\prime}$-$J$ model for the different sets of parameters:  
$t^{\prime}=-0.2t, t^{\prime\prime}=0.15t$  (solid curve),  
$t^{\prime}=-0.5t, t^{\prime\prime}=0.15t$ (dashed curve),
$t^{\prime}=-0.2t, t^{\prime\prime}=0$ (dotted-dashed curve).
Other parameters are: $t/J=2.5, t_S=-2t_N=-J/4$.
} 
\label{fig2}
\end{figure}

We also found  an important feature of the energy spectrum of the 
proposed model: if the values of $t^{\prime \prime}$, $t_S$, $t_N$ 
are not small (average and larger), the shape of the bandwidth is 
fully insensitive to the $t^{\prime}$ changes. Changes of $t^{\prime}$
only shift the energy of ground state and change the 
quasiparticle residue at the top of the band. Opposite to it, 
$t^{\prime \prime}$ strongly varies the $(\pi/2,0)$ position. These 
results are shown in Fig. \ref{fig2}. 
They are easily understood remembering that the higher energy states 
are unstable. When $t^{\prime \prime}$ and three-site terms are 
not very small, they already form the band up to the characteristic 
energy $2J$, and further changes of $t^{\prime}$ (even in a broad 
region) touch only the higher states, which are unstable.
It is interesting that the further increasing of $t^{\prime\prime}$ 
($>0.15t$) also do not change the shape of the band.

Summarizing, we showed that the generalized version of the 
$t$-$t^{\prime}$-$J$ model  accurately derived from the three-band
 model describe very well the experimental results of ARPES on 
Sr$_2$CuO$_2$Cl$_2$ system \cite{Wells}. Parameters of the model, at 
which good agreement achieved, are from realistic regions \cite{Bel1} 
and so they are strongly justified. Hopping integral to the 
next-next-nearest neighbor site ($t^{\prime \prime}$) as well as 
three-site hopping terms ($t_S$, $t_N$) is found to be the key 
parameters for the description of the flat region along  
$(0,0)-(\pi,0)$ line.  It is argued that the isotropy of the spectra 
around the minimum $(\pi/2,\pi/2)$ easily arise at any (not very 
small) $t^{\prime}$, $t^{\prime \prime}$, $t_S$, $t_N$ parameters  of 
the definite sign.  In addition we found that the shape of the 
spectrum is insensitive to varying of $t^{\prime}$ if the other 
parameters are not small. Thus, the model has
 some rigidity to 
parameter changes. Smaller intensity of the photoemission peaks at 
the top of the band can be directly related to the small 
quasiparticle residues at these points.

We thank O. Starykh for helpful discussion. This work was supported 
in part by the  Russian Foundation for 
Fundamental Researches, Grant No 94-02-03235.
V. B. is supported by the Portugal program PRAXIS XXI /BCC/ 4381 / 94.
A.C. is supported by the research 
program of International Center for Fundamental Physics in Moscow.
V. S. acknowledges the  support by ISSEP, Grant No. s96-1436.

%***********************************************************************

\vskip .2cm
%\newpage
%***********************************************************************
{\Large \bf Figure captions} \\ \vskip 0.1cm
%***********************************************************************
%-----------------------------------------------------

%***********************************************************************

\end{document}